\documentclass[]{piparticle-final}
\usepackage{graphicx}
\usepackage{amsmath}

\begin{document}

\volume{1}               
\articlenumber{010008}   
\journalyear{2009}       
\editor{J. J. Niemela}   
\reviewers{H. Cabrera Morales (Inst. Venezolano Invest. Cient\'{\i}ficas, Venezuela)}  
\received{9 November 2009}     
\accepted{29 December 2009}   
\runningauthor{D. Kunik \itshape{et al.}}  
\doi{010008}         

\title{Surface Percolation and Growth. An alternative scheme for breaking the diffraction limit in optical patterning}

\author{D. Kunik,\cite{inst1, inst2}\thanks{E-mail: dkunik@df.uba.ar}  
        \hspace{1ex}L. I.  Pietrasanta,\cite{inst2,inst3} 
        \hspace{1ex}O. E. Mart\'{i}nez\cite{inst1, inst2}}

\pipabstract{
A nanopatterning scheme is presented by which the structure height can be controlled in the tens of nanometers range and the lateral resolution is a factor at least three times better than the point spread function of the writing beam. The method relies on the initiation of
the polymerization mediated by a very inefficient energy transfer from a fluorescent dye molecule after single photon absorption. 
The mechanism has the following distinctive steps: the dye adsorbs on  the substrate surface with a higher concentration than in the bulk, upon illumination it triggers the polymerization, then isolated islands develop and merge into a uniform structure (percolation), which subsequently grows until the illumination is interrupted.
This percolation mechanism has a threshold that introduces the needed nonlinearity for the fabrication of structures beyond the diffraction limit.}
\maketitle

\blfootnote{
\begin{theaffiliation}{99}
   \institution{inst1} Departamento de F\'{\i}sica, Universidad de Buenos Aires, Buenos Aires, Argentina.
   \institution{inst2} Consejo Nacional de Investigaciones Cient\'{\i}�ficas y T\'{e}cnicas, Argentina.
   \institution{inst3} Centro de Microscop\'{\i}as Avanzadas Facultad de Ciencias Exactas, Universidad de Buenos Aires, Buenos Aires, Argentina.
\end{theaffiliation}
}
\section{Introduction}

The development of new techniques for the fabrication of smaller and smaller structures has become an objective of great relevance for many fields in science and technology \cite{ref1,ref2}. This includes semiconductor industry \cite{ref2}, MEMS \cite{ref3, ref4}, biology [4--7]
, microfluidics \cite{ref8}, material science and technology, among the most demanding [2]. Techniques include optical lithography, scanning electron lithography \cite{ref1,ref2}, dip-pen patterning \cite{ref9}, magnetolithograpy \cite{ref10,ref11}, ion milling \cite{ref12}, and many others. Among these techniques, optical lithography is the most widely used due to its inherent simplicity, mature development, fast production speed and less stringent ambient requirements \cite{ref1, ref2}. The far field optical lithography, both in projection and scanning, has a first principle limitation in size reduction, the diffraction limit, as light cannot be focused below a size of about half the wavelength used. The main approach to circumvent this problem is to reduce the light wavelength and this is the roadmap traced by the semiconductor
industry \cite{ref13}, now using less than 100nm sources mostly from synchrotron radiation but also from other new developments such as short wavelength lasers \cite{ref13, ref14}. Another approach is to avoid the far field limit by near field approaches \cite{ref15}, but these techniques suffer from the same drawbacks of other sophisticated scanning methods, slow throughput and small scanning areas, unless very sophisticated tricks are developed \cite{ref16}. It is now well recognized that the diffraction limit is derived assuming a linear response of the media to the light and this can be a way to circumvent the limitation imposed by diffraction \cite{ref17}. This approach has been used in many microscopy techniques and also in optical lithography with the main advantage of allowing the construction of three dimensional structures, but the increase in
resolution was marginal due to the need of longer optical wavelengths to target the same material transitions with more photons [17--19]
. In the last decade, new microscopic techniques with super-resolution have been developed, that rely on some complex nonlinearities, such as stimulated emission depletion (STED) \cite{ref20}. Recently, several researchers have reported ways to use these techniques using such nonlinear methods and switching strategies for photolithography [21--23]
.

In this work we present a new concept on how to circumvent the diffraction limit in optical lithography by placing the nonlinearity not in the light-matter interaction but in the material growth mechanism itself. The technique uses a dye molecule that initiates the
polymerization reaction from an excited state with a very low efficiency. The mechanism relies on the following distinctive steps:
\begin{itemize}
\item[(1)]
 	Mixture of the dye molecules and polymer in adequate proportions. 
\item[(2)]
 Deposition of a drop of the mixture on a transparent substrate. 
\item[(3)]
	 Adsorption of a fraction of the dye molecules on the substrate material. 
\item[(4)]
 Illumination of the mixture with a focused beam through the substrate at the absorption band of the dye. 
\item[(5)]
 Initiation of the polymerization by due to a very low quantum yield energy transfer from the dye to the polymer. 
\item[(6)] 
Percolation of the structure growing at the substrate surface. 
\item[(7)]
Deposition of fresh dye molecules from the bulk onto the growing surface.
\item[(8)]
Photoinitiation of the fresh molecules as the structure grows.
\end{itemize}
We will show how this steps allow the controlled growth of the structures with 10nm resolution in vertical direction (direction of propagation of the light), and that the lateral resolution can be increased by a factor of at least three, as compared to the diffraction
limit.

\section{Materials and Methods}

\begin{figure}
\begin{center}
\includegraphics[width=0.48\textwidth]{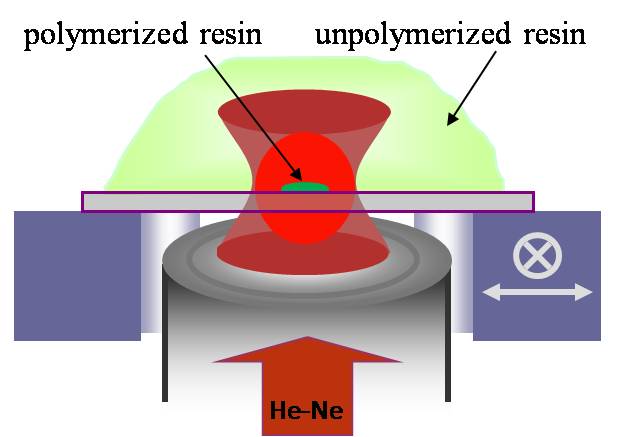}
\end{center}
\caption{Experimental setup.} \label{figure1}
\end{figure}

The experimental setup is shown in Fig. \ref{figure1} and it is similar to that used in previous works \cite{ref7, ref24, ref25}. The main difference is that the photo curable resin was not triggered by exciting the UV initiator by two-photon absorption directly but instead, the energy transfer to activate the polymerization was mediated by a dye molecule excited by one photon
excitation as described in \cite{ref24}. Different blends of dyes and resins were prepared and a drop of the blend was deposited onto a cover slip positioned in an inverted microscope setup equipped with a motorized stage. The laser beam was focused onto the sample by means of a high numerical aperture air objective (UplanSapo 40x, NA=0.9, Olympus, Tokyo, Japan). Control of the focus was made by means of a CCD camera imaging the back reflection of the laser beam, and the sample focus was adjusted in order to minimize the image size in the CCD camera \cite{ref25}. Because the resin had a refractive index matching that of the glass cover slip used as substrate, the surface had to be focused before the blend drop was placed. The
choice of the laser to excite the dye that triggers the polymerization process, depends on the absorption spectrum of the dye used. The results shown in this work correspond to cyanine (ADS675MT, American Dye Source, Quebec, Canada) and oxazine dyes (Nile-Blue 690 Perchlorate, Exciton, Ohio, USA) excited with a He-Ne laser emitting at 632.8 nm but similar results were obtained with different infrared cyanines dyes (ADS740, ADS760 and ADS775PI, American Dye Source, Quebec, Canada and HITC Iodide and LDS821, Exciton, Ohio, USA) excited with a Ti:Sapphire laser in cw mode tuned to match the absorption spectrum of these dyes. The laser power was adjusted by inserting neutral density filters in the beam path, and the exposure time was controlled with the scanning speed. In order to draw the desired pattern, a shutter was used to turn on and off the illumination while the scanning was being made. Tapping-mode AFM was performed in dry nitrogen using a NanoScope IIIa Multimode-AFM (Digital Instruments-Veeco Metrology, Santa Barbara, CA, USA) and images were acquired simultaneously with the height and the phase signals. Images were processed by flattening, using NanoScope software to remove background slope.

Different polymer-dye blends were prepared changing both the UV curing adhesive (NOA 60, NOA 63 or NOA65, Norland Products Cranbury, NJ, USA) and the dyes used. To ensure that the UV photo-sensitizer played no role in the process, a special batch of NOA60 resin without the UV sensitive additive was provided by the manufacturer, (Norland Products) which yielded similar results. The following steps were taken: (1) Select an adequate cosolvent for both the dye and the adhesive, in most cases ethanol, methanol and acetone worked well. We used methanol for our reported essays. (2) Make a concentrated solution of the dye in the solvent, typically 10 mM in methanol. (3) Add the dye solution to the polymer resin up to the desired concentration. (4) Place a coverslip on the microscope and set the focus at the surface. (5) Place a drop of blend on the coverslip. (6) Scan the laser to draw the desired structure. (7) Rinse the coverslip immersing it in ethanol and acetone. A final rinse of a few seconds in methylene chloride is helpful to efficiently remove the remanent unpolymerized resin.
\begin{figure}
\begin{center}
\includegraphics[width=0.48\textwidth]{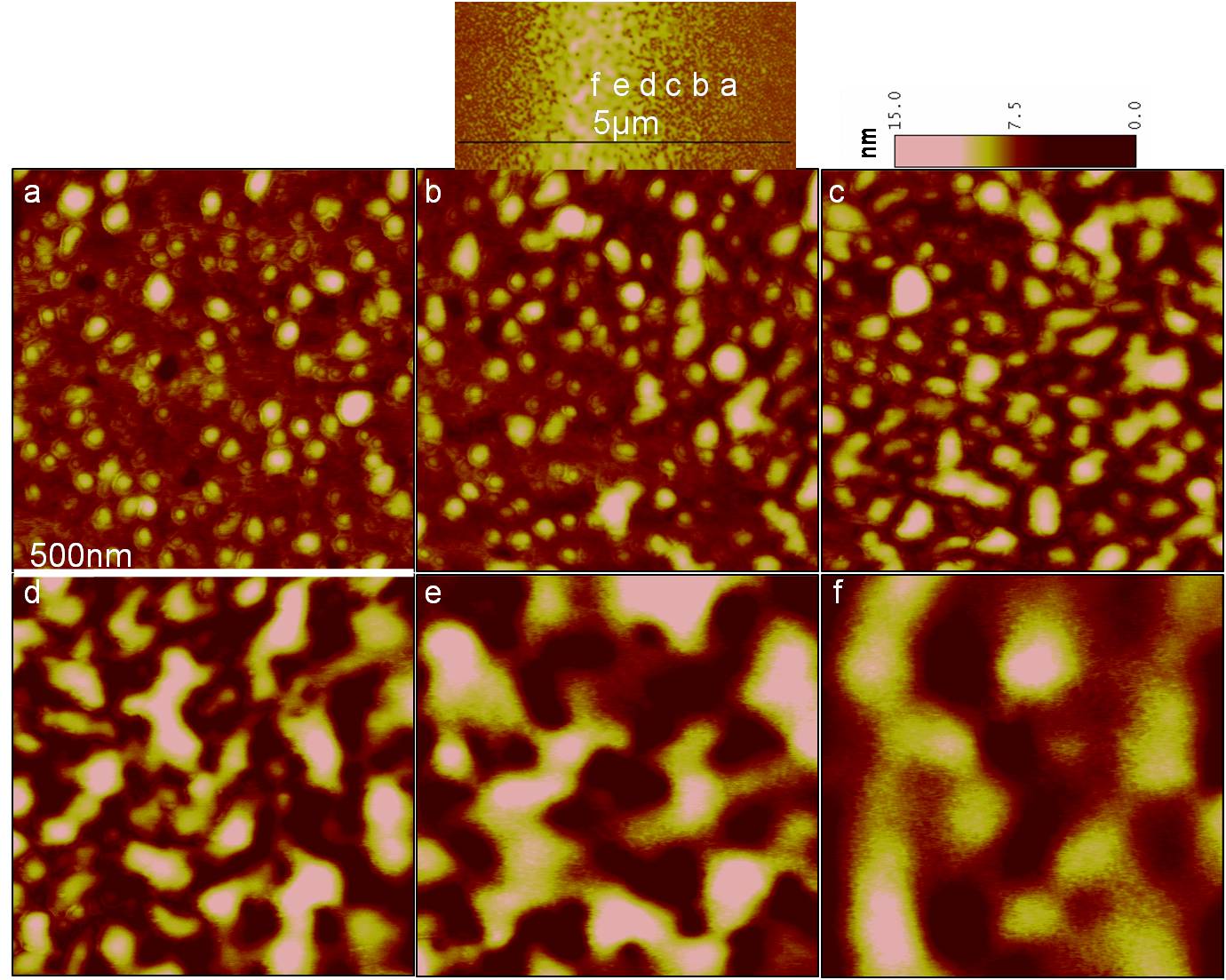}
\end{center}
\caption{Different stages of the surface percolation process. Upper inset is an AFM image of a wide line showing the location of the subsequent detailed images. AFM height images taken from the periphery (a) toward the centre of the structure (f). At the periphery small clusters are observed. Towards the center of the structure (higher laser intensity), the size of the clusters grow until percolation takes place. The size of each image was 500 nm $\times$ 500 nm and the displacement between adjacent images was approximately 400 nm.} \label{figure2}
\end{figure}

\section{Results}

In order to study the different stages of the process, we fabricated different samples under specially designed conditions. After selecting the polymer and the dye, three parameters remain for the control of the size of the structure. 
\begin{itemize}
\item[(a)]
	The dye concentration
\item[(b)]
	The beam intensity
\item[(c)]
	The light exposure time (or scan velocity).
\end{itemize}

\begin{figure}
\begin{center}
\includegraphics[width=0.48\textwidth]{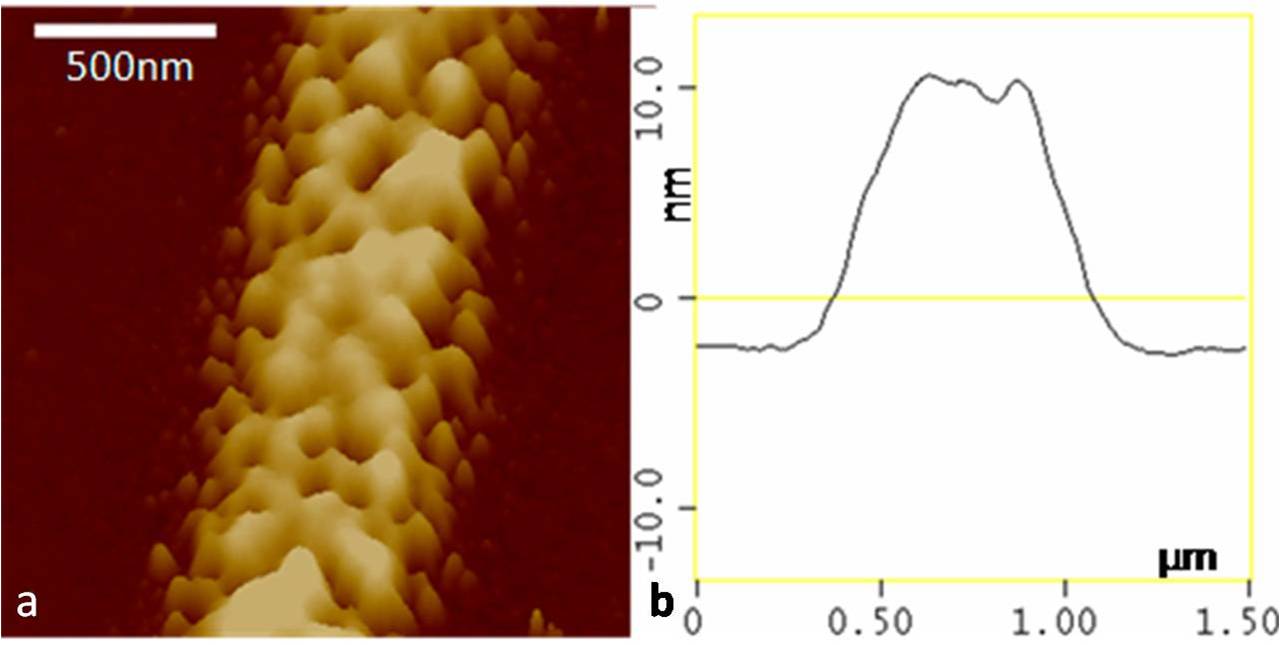}
\end{center}
\caption{Structure at the percolation threshold showing full coverage. (a) 3D reconstruction of a
AFM topographic image. (b) average cross section.} \label{figure3}
\end{figure}

Figure \ref{figure2} shows AFM images of a wide structure fabricated with a mixture of the optical adhesive NOA 63 and the laser dye ADS675MT. The dye conentration was 1 mM. The sample was illuminated with a He-Ne laser beam and the power at the sample was 1 mW. The laser beam was defocused in order to produce a broad line about 5 $\mu$m wide in such a way that in a cross section the different stages of the polymerization process can be visualized. The scan speed was 10 $\mu$m $\rm{s}^{-1}$ and the different degree of coverage can be
detected as one moves towards the line center (maximum intensity). The structure was detailedly scanned in six different regions going from the periphery of the structure (a) towards the center of the line (f). The different scans are 500 nm $\times$ 500 nm in size and located at 400 nm steps from each other. The first scan shows very isolated polymer islands of different sizes with heights from a few nm to about 7 nm. As the center is approached, the size of the islands grows, but the height only grows marginally, showing a percolation process as more dye molecules located at the surface start the polymerization process at different locations. Towards the center, the percolation phenomenon becomes evident, with islands merging towards a larger and larger coverage, but with only a minor increase in the height. With an increase in the intensity by an improvement in the focusing of the beam, a total coverage is obtained, as shown in Fig. \ref{figure3}, still maintaining 13 nm 
in height. The laser power was 35 $\mu$W, the scanning velocity was 16 $\mu$m $\rm{s}^{-1}$ and the blend used was a mix of the optical adhesive NOA60 and the dye ADS675MT at a concentration of 1 mM.

Once the percolation threshold is reached, the structure grows slowly, as more dye molecules are adsorbed in the fresh polymer surface. This continues until the illumination is turned off or the beam walks away from that portion of the surface. Different concentrations
and speeds were used in order to fabricate lines of different heights and widths. The concentrations used were 1 mM, 500 $\mu$M, 100 $\mu$M, 50 $\mu$M, and 10 $\mu$M of a NOA60-ADS675MT blend. For each dye concentration, AFM topographies in tapping mode and phase images were obtained for lines manufactured at different
speeds with the same beam intensity in order to cover a wide range of stages of the percolation and growth process. The phase images (not shown here) helped to easily distinguish the polymer from the substrate due to their different hardness. When the volume dye concentration is very large (1 mM and 500 $\mu$M) the surface percolation phenomenon can be observed (as shown already for 1 mM) but the growth was hindered by the fact that at these high concentrations a volume reaction takes place. This fact is  due to the existence of a second threshold fluency that gives rise to a volume percolation, as will be discussed in the next section.

\begin{figure}
\begin{center}
\includegraphics[width=0.48\textwidth]{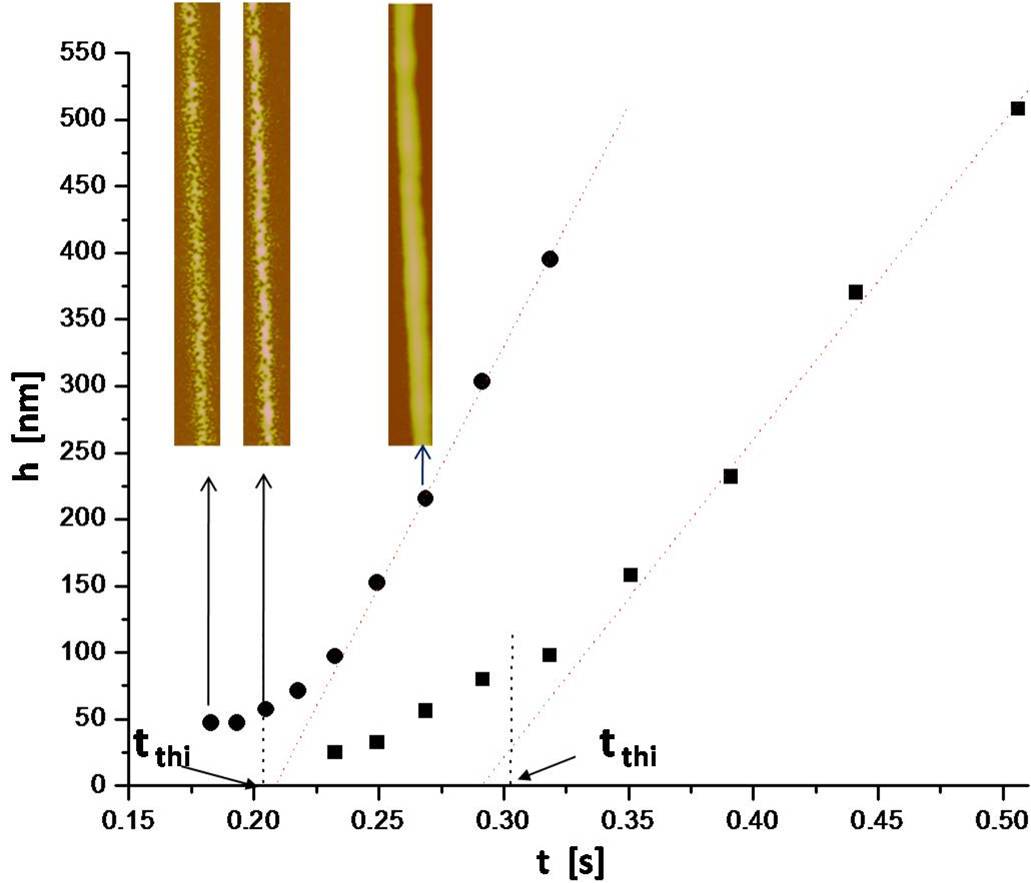}
\end{center}
\caption{Height of the lines as a function of the exposition time for samples with dye
concentration of 50 $\mu$M (squares) and 100 $\mu$M (circles). Examples of structures (topographic
AFM images) below, at and above the percolation threshold are also shown.} \label{figure4}
\end{figure}

In Fig. \ref{figure4}, a plot of the maximum height of the lines as a function of the exposition time (defined as the ratio between the beam size and the scan speed) is presented for the samples with dye concentrations 100 $\mu$M (squares) and 50 $\mu$M (circles). The beam power at the sample was 290 $\mu$W
(100 $\mu$M) and 460 $\mu$W (50 $\mu$M) and the beam diameter, Full Width Half Maximum (FWHM), was 800 nm. For the sample made with 100 $\mu$M, the speed was varied from
line to line in steps of 0.2 $\mu$m $\rm{s}^{-1}$, from 2.2 $\mu$m $\rm{s}^{-1}$ for the slower scan (higher line) to 3.8 $\mu$m $\rm{s}^{-1}$ for the fastest one (lower
line). While for the sample made with 50 $\mu$M the range of the speed was 1.4 $\mu$m $\rm{s}^{-1}$ to 3 $\mu$m $\rm{s}^{-1}$. From the AFM topographies, the surface percolation threshold time $t_{thi}$ (threshold time obtained from the image) was determined and is indicated in the plot. We also show three characteristic lines at both sides of the percolation threshold of the 100 $\mu$M sample.

As the scanning speed is reduced, the exposure time increases and the structures generated show a transition from isolated islands to full coverage and afterwards, a gradual growth in height. A similar experiment was performed for the five concentrations indicated before. However, for the 10 $\mu$M sample, the percolation threshold was not reached. For the samples with concentrations of 1 mM and 500 $\mu$M after the surface percolation was reached, the structure did not grow gradually but instead grew suddenly from less than 20 nm to 1 $\mu$m or more. The experimental data presented in Fig. \ref{figure4} show a clear nonlinear growth mechanism with an apparent linear asymptotic behavior for large exposure times. It can also be observed that the linear asymptote crosses the time axis at approximately the threshold exposure time.

\begin{figure}
\begin{center}
\includegraphics[width=0.48\textwidth]{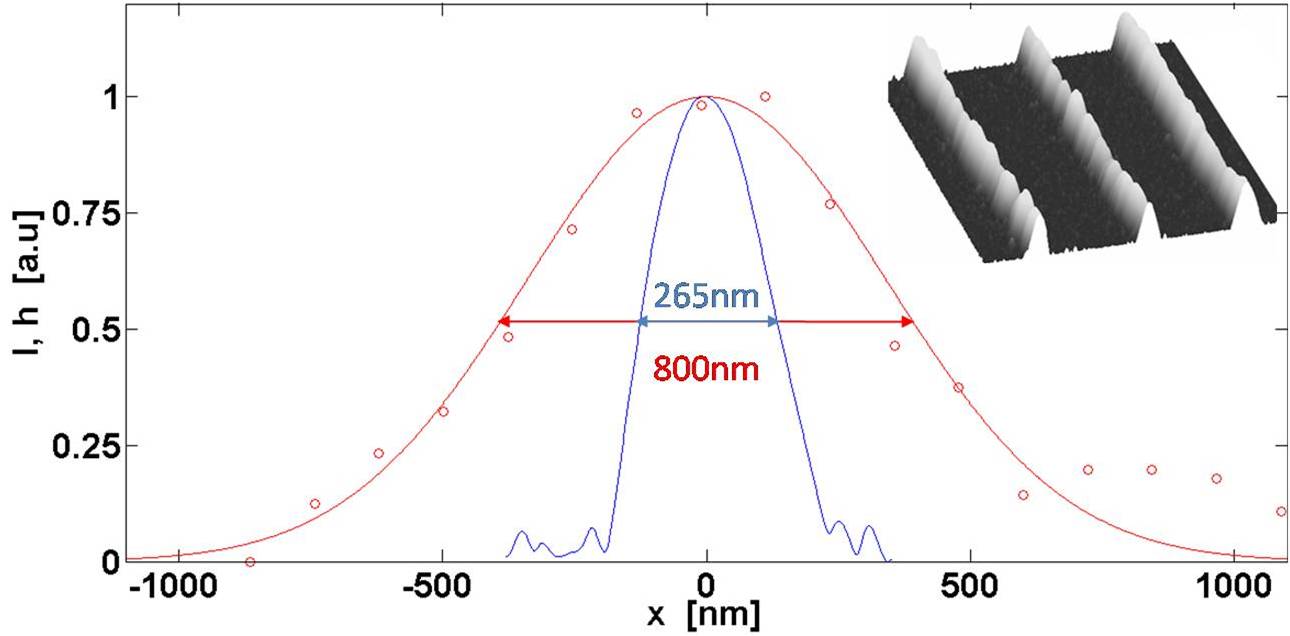}
\end{center}
\caption{Subdriffaction structures made with the blend NOA63-Nile Blue 1 mM.
Normalized PSF of the laser beam (red) and AFM profile of the structure (blue). Inset: 3D reconstruction of the subdiffraction structures topography.} \label{figure5}
\end{figure}

Figure \ref{figure5} shows the lateral reduction below the diffraction limit of structures fabricated with 
Nile Blue 690 perchlorate and NOA63. The Point Spread Function (PSF) is plotted (red) together with a line profile of the structures (blue). A 3D topographic reconstruction of the structures is also shown as inset. The height of the structures was 67 nm and the width (FWHM) was 265 nm. The PSF of the beam was determined by measuring the scattered light from a gold nanoparticle 80 nm in diameter and the width was 800 nm (FWHM). The lateral resolution is at least three times better than the beam size, considering that the tip shape was not deconvolved in the measurement. This reduction below the diffraction limit is a clear evidence of the nonlinear growth mechanism also shown in Fig. \ref{figure4}. Similar results were obtained for many combinations of dye molecules and UV curing resins. The common denominator of all the dyes used, which included cyanines and oxazines, is that they are all good laser dyes (meaning high fluorescence efficiency, low triplet formation, low excited state absorption). Experiments with methylene blue were not successful (no polymerization was observed) even when the dye was bleached. Another common aspect is that all the dyes used have the chromophore positively charged and they were all very poorly soluble in the polymer
blend.

\section{Discussion}

As shown in the previous section, the process has two very distinctive stages. A first stage in which the polymerization process takes place mainly at the substrate surface with the formation of isolated islands that gradually merge, and a second stage after the surface percolation in which the structure grows gradually until the illumination is turned off. As only marginal growth is detected during the percolation process. We will discuss two models in order to explain the two stages separately.

\subsection{Percolation model}

As the smallest islands at the very beginning of the process are too small to be measured with our AFM microscope (24 nm radius tip) we cannot accurately determine the island size distribution. Hence, we will only model this stage semi-quantitatively and we will show that the surface percolation process requires a large adsorption of dye molecules at the substrate surface. If this is not the case, or else if the volume concentration is too large, a volume percolation precedes the surface percolation and a thick structure of the size of the illuminated volume is generated. In order to compare these two situations, we modeled the percolation process as follows: 
\begin{itemize}
\item[(a)]
We assumed that the dye molecules were distributed in an ordered lattice with a distance between neighbors given by the concentration used. 
\item[(b)] 
At the substrate surface, a similar lattice was assumed but with a different neighbor distance kept as a parameter. 
\end{itemize}

\begin{figure}
\begin{center}
\includegraphics[width=0.48\textwidth]{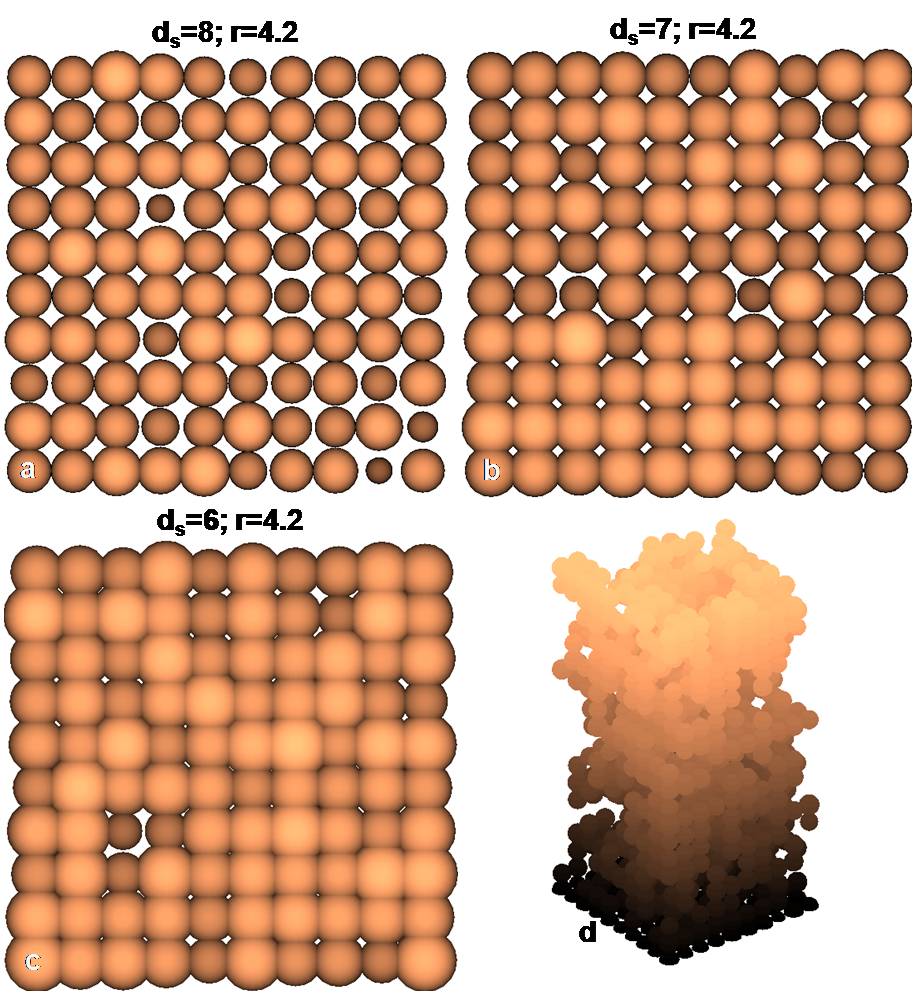}
\end{center}
\caption{Simulations of the percolation process. Spheres with average radius r are randomly created. (a), (b) and (c) show structures simulated as the distance between neighbors {$ d_s $}   was diminished. (d) shows the volume percolation for the case where the nearest neighbor distance in bulk equals that at the surface.} \label{figure6}
\end{figure}

Despite the fact that in the experiment the dye molecules are actually randomly distributed, we found that adding this to the model did not modify the result. The result being that at equal nearest neighbor distance the volume percolation precedes the surface percolation. It was assumed that each dye molecule (lattice site) could initiate the polymerization process at random with a probability that increases with the illumination time and intensity. Each polymerization triggered generated a sphere of polymer with a Gaussian distribution in size. We found no significant differences in the results by changing the variance of the distribution. After a given time, the process is stopped, each sphere is linked with any neighboring sphere if their distance is smaller that the sum of the radii and only the spheres that are linked (directly or through other neighbors) to the substrate are kept (sample rinsing). The results of some typical situations are shown in Fig. \ref{figure6}. For very low concentrations or light doses, 
isolated islands appear at the surface because the structures generated in the bulk do not touch each other and are washed. As the surface concentration is increased, a more densely packed structure develops, as shown in the sequence of Fig. \ref{figure6} (a) to (c). In this last case, the surface is fully covered by the polymer spheres. If the surface concentration is kept with a nearest neighbor distance larger than about one half of the bulk average distance, the surface percolation never occurs before a massive reaction takes place due to the percolation of the structure in bulk. This is shown in the simulation of Fig. \ref{figure6}(d) where the same distance between neighbors was used for the surface and bulk. In brief, the surface percolation process as modeled yields structures with typical sizes that depend on the nearest neighbor distance, and a much lower distance (high surface concentration) is required if the surface phenomenon is to prevail the bulk process. This result, together with the fact that at the percolation threshold lower lines were obtained with high concentrations, indicate that the substrate surface was not saturated and fully covered by the dye molecules. The surface dyes concentration grows with the volume concentration. This also explains why at higher concentrations (1 mM and 500 $\mu$M) no growth could be observed but instead a sudden bulk polymerization at the entire illuminated volume. Also that for the low concentration case (10 $\mu$M) the surface concentration was too low and hence the large distance between neighbors (larger than the size of the isolated islands) did not allow the percolation process to take place.

\subsection{Growth model}

\begin{figure}
\begin{center}
\includegraphics[width=0.48\textwidth]{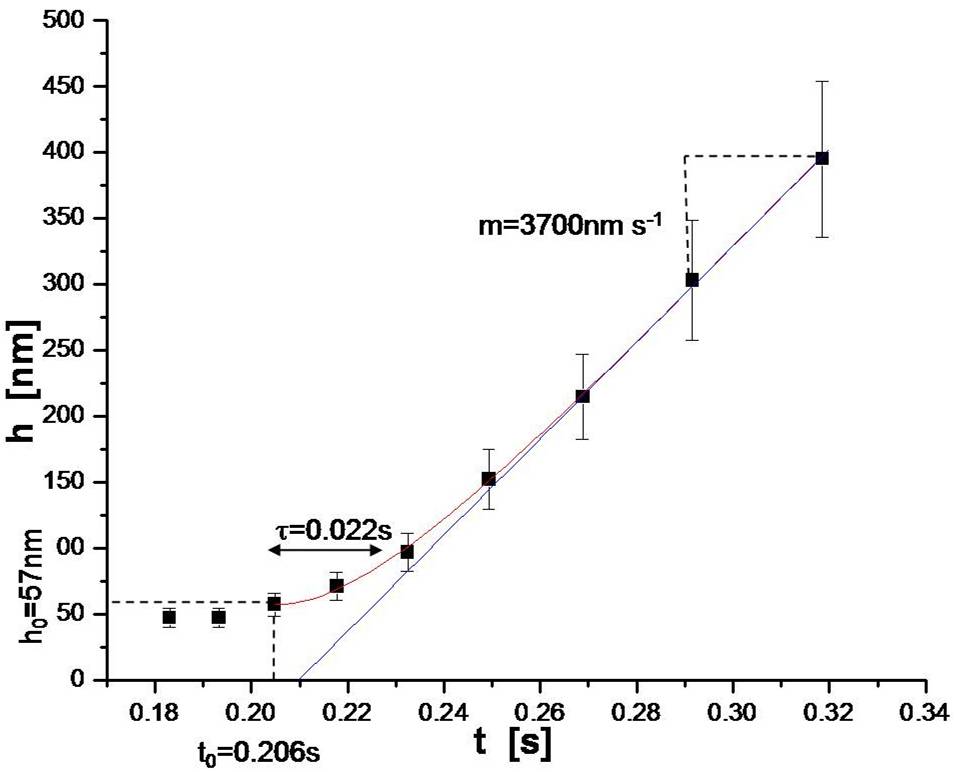}
\end{center}
\caption{Fit of the data for the sample made at 100 $\mu$M dye concentration. } \label{figure7}
\end{figure}

Once the surface is fully covered, the growth of the structure requires the adsorption of fresh dye molecules at the surface of the growing polymer structure. We will model this situation by assuming, in a simple manner, that the growth rate follows the rate equation	

\begin{align}
\frac{dh}{dt}
= \nu \Phi_{p}  I \sigma \rho . \label{eq1}
\end{align}
Where $I$ is the beam intensity, $\sigma$ is the absorption cross section of the dye molecule, $\rho$ is the surface dye concentration, $\Phi_p$ is the efficiency of the initiation of the polymerization process (the inverse of the number of photons required to trigger one polymerization reaction) and $\nu$ is a characteristic volume indicating the size of the structure created once a polymerization event is triggered. This simple model is consistent with the assumption that each molecule can trigger only one event and yields a typical size of the polymer structure or, alternatively, that each molecule can catalyze more than one event and the size is proportional to the number of polymer chains initiated. To complete the description, an equation is needed for the dynamics of the surface density $\rho$. Such equation should take into account the adsorption-desorption process and the dye consumption due to the reaction and it can be written as

\begin{align}
{\frac{d\rho }{dt}} =  - \kappa \rho + \alpha C(\rho _0  - \rho ) - \Phi _p I\sigma \rho . \label{eq2}
\end{align}
Where $\kappa$ is the desorption rate, $\alpha$ is the adsorption or sticking coefficient proportional to the number of collisions one molecule has with the polymer surface and the probability of sticking to that surface, $C$ is the volume concentration to be assumed constant, $\rho_0$ is the surface density of sites for the dye molecule (being $\rho_0 - \rho$ the density of available sites). The three components in the right hand side of the equation account for the desorption rate, adsorption rate and dye consumption respectively. Any gradient in the volume concentration $C$ is neglected for the sake of simplicity and is a good approximation if the rate is dominated by the dye consumption term, as the structure would be growing faster than the development of the concentration gradient.

The solution to Eq. (2) is

\begin{align}
\rho (t) = \rho _e \left( {1 - e^{ - \,\frac{t}{\tau }} } \right), \label{eq3}
\end{align}
were

\begin{align}
\rho _e  = \frac{{\rho _0 \,\alpha \,C\,}}{{\alpha \,C + \kappa  + \sigma I\Phi _p }} \label{eq4}
\end{align}
is the equilibrium surface concentration and

\begin{align}
\tau  = \frac{1}{{\alpha \,C + \kappa  + \sigma I\Phi _p }} \label{eq5}
\end{align}
is a characteristic transition time.

\begin{figure}
\begin{center}
\includegraphics[width=0.48\textwidth]{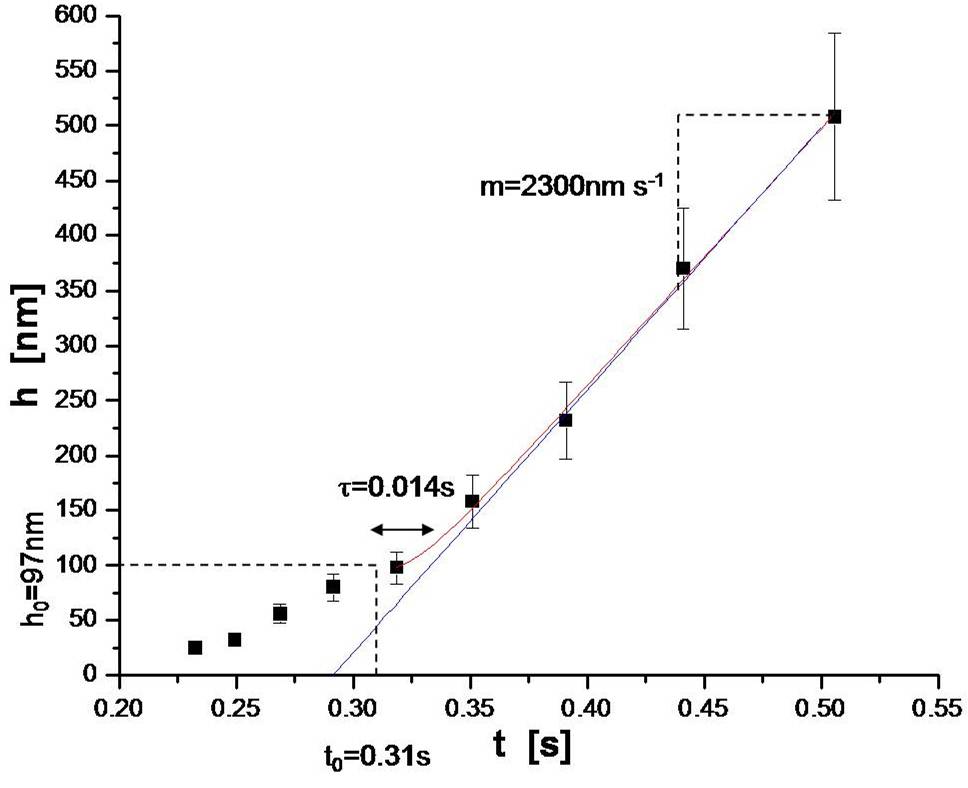}
\end{center}
\caption{Fit of the data for the sample made at 50 $\mu$M dye concentration. } \label{figure8}
\end{figure}
Inserting Eq. (3) in Eq. (1), the solution for the height evolution is
\begin{align}
h(t) = m\,(t - t_0 ) + h_0  - m\,\tau \,\left( {1 - e^{ - \,\frac{{t - t_0 }}{\tau }} } \right); \label{eq6}
\end{align}
were
\begin{align}
m = \frac{{v\,\alpha \,C\,\rho _0 \,\sigma \,I\,\Phi _p }}{{\alpha \,C + \kappa  + \sigma \,I\,\Phi _p }}, \label{eq7}
\end{align}
$t_0$ is the percolation threshold time delay (initial time for the growth equation) and $h_0$ is the height of the structure at the percolation threshold.  Equation (6) shows an initial nonlinear growth towards an asymptotic linear growth with slope $m$. The characteristic transition time towards the asymptotic behavior is precisely the parameter $\tau$ defined in Eq. (5). 
The results shown in Fig. \ref{figure4} were fit using Eq. (6) and the result is shown in Fig. \ref{figure7} for the 100 $\mu$M dye concentration sample and in Fig. \ref{figure8} for the 50 $\mu$M dye concentration sample. From these fits, the following conclusions can be drawn: 
\begin{itemize}
\item[(a)]
the linear asymptote is evident in both cases and extrapolates crossing the time axis at the measured percolation time
\item[(b)] 
the slope of the asymptote grows with the dye concentration. 
\item[(c)]
The surface percolation height decays with the concentration ranging from almost 100 nm for the 50 $\mu$M concentration, to 13 nm for the 1 mM case. 
\end{itemize}

The last result is consistent with the fact that the surface is not saturated with the dye but instead the surface concentration grows with the bulk concentration. By examining the equations obtained for the growth rate, these results are consistent with having the surface concentration rate dominated by the dye consumption

\begin{align}
\sigma I \Phi _p  \gg \kappa + \alpha C, \label{eq8}
\end{align}
which yields a slope proportional to the concentration as from Eq. (7) in this limit

\begin{align}
m = v\,\alpha \,C\,\rho _0.  \label{eq9}
\end{align}
And in the same limit Eq. (5) yields

\begin{align}
\tau  = \frac{1}{{\sigma \,I\,\Phi _p }}. \label{eq10}
\end{align}

The ratio of the slopes between the 100 $\mu$M and the 50 $\mu$M is $ \frac{m_{100}} {m_{50}}=1.6 \pm 0.3 $ which is, within the experimental error, the ratio between concentrations $\frac{C_{100}}{C_{50}}=2 \pm 0.4$ . The experiment for the two concentrations was done at different beam powers yielding $\frac{\tau_{100}}{\tau_{50}}=1.43 \pm 0.9$
and $\frac{I_{50}}{I_{100}}=1.58\pm 0.0.3$ which show an excellent agreement with this limiting approximation. From Eq. (10) and the characteristic time obtained by fitting the experimental results, we found the efficient of initio of the polymerization process is

\begin{align}
\Phi _p  = \frac{1}{{\sigma \,I\,\tau }} = (7 \pm 2)\,10^{ - 7} \label{eq11}
\end{align}

\section{Conclusions}
A technique has been described that allows the fabrication of polymer structures by scanning a laser beam and yielding features with sizes below the diffraction limit. The structure height can be controlled with 10 nm resolution in the range from a few tens of nanometer to hundreds of nanometers. The width was shown to be reduced by at least a factor of three times the diffraction limit.

The technique relies on the use of a light sensitive dye that transfers the absorbed energy to the resin and triggers the polymerization process. The exact mechanism could not be clearly separated as the reaction efficiency was shown to be extremely low (below 1ppm). This low yield appears to be crucial for the success of the technique. Our results show that the surface dye adsorption and the low dye polymerization efficiency are the keys to allow a smooth growth from surfaces. The surface dye concentration increases the speed of the polymerization at the surface with respect to the volume. On the other hand, the low yield of the triggered polymerization by the excited dye allows this effect to be observable. If the yield of the dye polymerization was high, no matter if there are more initiators at the surface than in the polymer volume, all excited dye would trigger a polymerization event and hence the whole volume illuminated would be polymerized. If the polymerization speed was uniform, the surface growth would take place at a lower speed than the volume growth due to the higher neighbors available in volume, and in this case, the polymerization would also occur in the whole illuminated volume. As was shown, two distinct processes take place in an almost sequential manner. The first one is a surface nucleation of isolated islands that finally percolate into a uniform surface coverage with a characteristic size that decreases as the dye concentration increases. This is followed by a structure growth, as fresh dye molecules are deposited on the growing polymer surface. The process is stopped when the illumination is interrupted by shutting down the laser beam or moving it to a different spot where the process starts again. The percolation mechanism was modeled numerically and the simulations showed that a large surface adsorption of the dye molecules is needed in order to avoid the volume percolation and allow a surface growth mechanism. A model developed for the growth process permitted the fit of the experimental data and the determination of the extremely low transfer efficiency from photons to the polymer. How the growth rate increases with the dye concentration after an incubation time that is inversely proportional to the light intensity, was also explained.

\begin{acknowledgements}
We thank Norland-Products for the sample without photoinitiator. Grants from Universidad de Buenos Aires, CONICET (Argentina) and ANPCYT (Argentina) are acknowledged.
\end{acknowledgements}

\end{document}